\documentclass{osa-article}

 \journal{ome} 

\articletype{Research Article}

\usepackage{lineno}
\nolinenumbers

\begin{document}

\title{Dual-Band Metal-Insulator-Metasurface Absorber}
\author{Ibrahim Issah,\authormark{1} Carlos Rodríguez-Fernández,\authormark{1} Mohsin Habib,\authormark{1} Semyon Chervinskii,\authormark{1}and Humeyra Caglayan\authormark{1}}

\address{\authormark{1}Faculty of Engineering and Natural Science, Photonics, Tampere University, 33720 Tampere, Finland.}

\email{\authormark{*}humeyra.caglayan@tuni.fi} 



\begin{abstract}
We explored a metal-insulator-metasurface structure exhibiting dual-band absorption in the visible and infrared regions with almost perfect absorbance. We demonstrated that the existence of the dual peak absorbance is due to the excitation of propagating surface plasmon and localized surface plasmon mode. We further utilized the excited degenerate propagating surface plasmon to enhance the fluorescence emission of an emitter located on top of the metasurface. This was accomplished by engineering a metasurface that excites a degenerate mode with resonance absorption bands close to the emission wavelength of fluorescent material. sThis condition helps to increase the rate of excitation and emission of an emitter by generating additional electron configurations when coupled to such metal-insulator-metasurface structures. This approach offers relevant potential in radiative engineering techniques.
\end{abstract}
\section{Introduction}
Metal-insulator-metal structures are artificially engineered multi-layered structures with an optical response that is dependent on their geometrical and structural properties\cite{Mou20201, https://doi.org/10.1002/adma.201200674}. Many interesting optical phenomena have been shown based on their unique geometrical and structural properties such as selective color filtering \cite{doi:10.1021/nl900755b, https://doi.org/10.1002/adom.201800851, doi:10.1021/acsami.1c15616}, absorption \cite{Luo:16, Aalizadeh2018, Vasi__2021}, plasmon hybridization \cite{doi:10.1063/1.5115818, https://doi.org/10.1002/adom.202000609,doi:10.1021/acsphotonics.9b00977}, surface-enhanced Raman scattering (SERS) \cite{doi:10.1021/nn901826q, Xu2017HybridMS}, multiplexing detector arrays \cite{doi:10.1021/nn3026468}, and sensing \cite{FENG2020103272, doi:10.1021/nl9041033, Khonina:21}. 
Among these, metamaterial absorbers (MAs) with a dielectric layer sandwiched between two metallic structures, create a resonance cavity that enhances its spectral absorption level to almost unity ($\approx$ 100\%) \cite{prakash_gupta_2021, He:13}. These absorption characteristics of MAs are linked to their structural configurations. By manipulating the geometrical parameters of these structures, one can attain unique absorption properties which are linked to the MAs; strong electric and magnetic resonances, reduction in the plasma frequency of metals, impedance matching, and the value adjustments in the permittivity and permeability of the MA structure \cite{abdulkarim2022review, doi:10.1142/S2010135X18500078}. 

In general, MAs are divided into two categories based on their applications (i.e., broad- and narrowband MAs) \cite{ma13225140, Issah:21, Shi2019a}. Broadband MAs exhibit absorption characteristics over a wide spectral range. They are investigated more thoroughly in literature due to their large absorption bandwidth in the infrared (IR) region \cite{Hle2012, Li:22, 10.3389/fphy.2022.893791}. As a result, broadband MAs are ideal for energy harvesting, especially in solar engineering and thermionic technologies. 
On the other hand, resonant MAs are narrowband absorbers that exhibit either single or multi- resonances with narrow spectral bandwidth and confined electromagnetic (EM) field characteristics. 
These confined EM fields excite different collective electron oscillations (plasmonic modes) in the MA structure that can be leveraged into radiative engineering.

\begin{figure}[!ht]
\includegraphics[width=\linewidth]{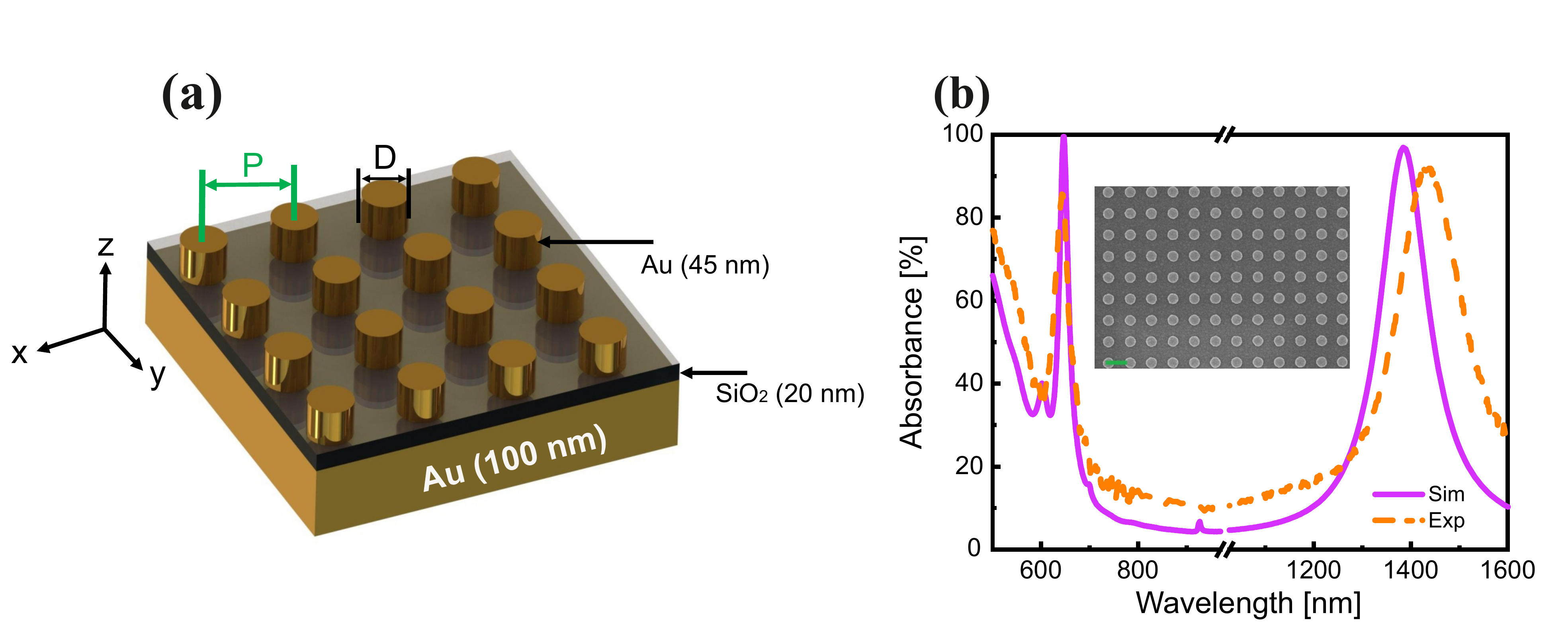}
\caption{\label{fig:fig1} (a) Schematics of the MIMS structure with Au nanoarray period $P$ and diameter $D$. (b) Simulated and experimental absorption spectra of the MIMS structure with a period of 600 nm and nanodisc diameter $D = 280nm$. The fabricated sample structural parameters: Au-film thickness  100 nm, SiO$_{2}$ spacer thickness (ST) 20 nm, and the Au nanodisc 45 nm with the SEM image shown as an inset (scale bar is 500 nm).}
\end{figure}

Plasmonic modes, in particular, can generate a strong local electric field $E$, that can enhance the excitation rate, $\gamma_{\text{exc}}$ ($\propto $ $|E|^{2}$) of a fluorescent material when coupled to it. Fluorescent materials (emitters) such as dye molecules, quantum dots, upconversion nanoparticles, conjugated polymers, and carbon dots, acts as excellent sensing probes \cite{JEONG2018102, Issah:22}. However, due to their emission drawbacks (i.e., photobleaching, autofluorescence, and low quantum yield) in fluorescence-based detection techniques, it is relevant to properly tailor photonic environments seen by these materials to enhance their fluorescence and photostability \cite{Prayakarao:19}. The presence of such an environment near a fluorescence material increases their rate of excitation and emissions by generating additional electron configurations of the fluorophore \cite{Aoki2008}. 


Recently, Ali et al. \cite{ma13225140} theoretically investigated a nano-patterned multi-resonant MA composed of Au-MgF$_{2}$-Au array structure operating between 600 and 950 nm wavelengths. They identified the absorption band in the visible and IR region with the excitation of two plasmonic modes (i.e., propagating surface plasmon (PSP) and localized surface plasmon (LSP) mode, respectively). Liu et al., \cite{doi:10.1021/nl9041033} also presented experimentally, a similar polarization-independent resonant MA with a focus on LSP mode excitation at $\lambda = 1.6 \mu m$. Although the absorption capability of different geometries and topologies of resonant MAs have been theoretically predicted, there has been only a limited number of reports on practically realizing such MAs \cite{Daniel2020, DANIEL2018,ma11030458} as well as extending its capabilities in plasmon-enhanced fluorescence techniques utilizing its strong electric and magnetic properties. 
Notwithstanding, emitters coupled with different photonic environments, although, exhibit enhanced fluorescence emission, yet, they suffer from quenching effects and photostability constraints. These also reduce the lifetime of the dye molecules due to non-radiative recombinations. As a result, many authors have strived in reducing these drawbacks of emitters coupled to different photonic environments, however, plasmonic modes in typical metallic structures' nonradiative effects limit the Purcell enhancement factor of emitters in the far-field \cite{Bozhevolnyi:16, JEONG2018102}.

As such, this work seeks to address these Purcell enhancement factor limitations of metallic structures by using a complementary metal-insulator-metasurface (MIMS) configuration capable of outcoupling nonradiative plasmonic modes and generating high field enhancements that can enhance emitters coupled to it. We designed, fabricated, and 
investigated the MIMS structure not only to identify its optical properties but also to unveil its fluorescence enhancement capabilities when coupled with a dye molecule embedded within a polymer matrix film. The MIMS structure is composed of a grating-like top periodic metallic pattern capable of outcoupling plasmonics modes which can extend the radiative emission of emitters in the far-field when coupled to it. Here, we leveraged on the PSP mode excitation of the proposed design and its enhancement capability when coupled to an emitter with an emission wavelength closed to the plasmonic mode spectral bandwidth.

\section{Results and Discussion}
\subsection{Structure design and spectral response of MIMS}
The MIMS structure in this study is composed of Au-SiO$_2$-Au nanodisc array stacked on SiO$_2$ substrate as shown in the schematic of Fig \ref{fig:fig1} (a). Dual-band absorption resonance is obtained [Fig. \ref{fig:fig1} (b)] in the visible and IR region for the following structural parameters: bottom Au-film thickness 100 nm, SiO$_2$ spacer thickness $ST$ 20 nm and nano-patterned Au array thickness 45 nm with a period ($P = $600 nm) and diameter ($D = $280 nm). The inset in Fig. \ref{fig:fig1} (b) depicts the scanning electron microscope (SEM) image of the fabricated MIMS structure. The visible absorbance peak at $\lambda_{1} = $ 647 nm has a spectral bandwidth of $\Delta \lambda \approx $20 nm with Q-factor $\approx 33$ as compared to the IR peak absorbance at $\lambda_{2} = $1390 nm  with a resonance Q-factor $\approx 13$.  
The absorbance $A$ of the MIMS structure is calculated using the relation $A$ = $1-R$  with almost zero transmittance ($T$) due to the high thickness of the bottom Au layer that exceeds its skin depth. Thus, to achieve perfect absorbance (A $\approx 1$), it is relevant to minimize the reflectance ($R$) of the MIMS structure through impedance matching with the surrounding medium and simultaneously eliminating the transmittance ($T$) by maximizing the structure losses\cite{Martinez2013, PhysRevLett.100.207402, doi:10.1021/nl9041033}.

To comprehend the origin of the dual-band resonance, we first examine the field distributions at the absorption wavelengths of $\lambda_{1}$ and $\lambda_{2}$, respectively. We calculated the electric field $|E|$ distribution of the MIMS structure in the x-y plane at the visible and the infrared (IR) spectral region (not shown here). We observed the existence of a quadrupolar mode (i.e., second-order LSP mode) in the visible spectral region and a dipolar mode in the IR region due to the collective excitation of conduction electrons in the Au-nanoarray. 

\begin{figure*}[!ht]
\includegraphics[width=\linewidth]{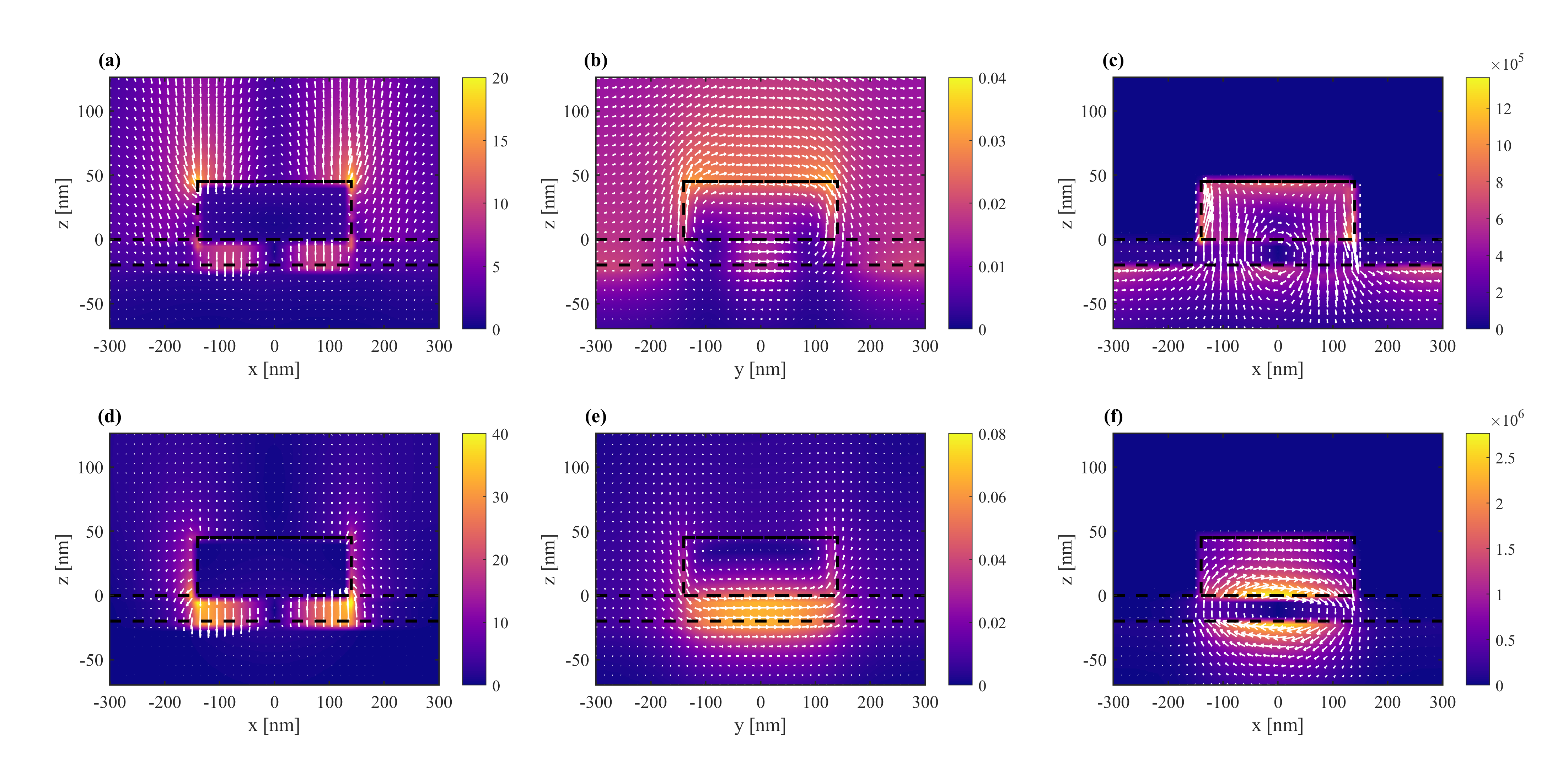}
\caption{\label{fig:fig4b} (a), (b), and (c) Electromagnetic field distribution at $\lambda_{1} \approx$ 647 nm. Similar field distribution at $\lambda_{2} \approx$ 1390 nm is presented in (d), (e), and (f). (a) and (d) Electric field |E|, (b) and (e) Magnetic field |H|, (c) and (f) Current density $J$ of the MIMS structure. Superimposed white arrows illustrate the vector field distributions and the black lines illustrate the unit cell x-z and y-z cross-sections.}
\end{figure*}

Furthermore, the electric field |E|, magnetic field |H|, and the current density (J) is computed in the x-z and y-z cross-section at $y = 0$ and $x = 0$ of the MIMS structure, respectively, as shown in Fig. \ref{fig:fig4b}, 
to additionally characterize the PSP and LSP spectral band of the MIMS structure in the visible and IR spectral region. Figure \ref{fig:fig4b} (a), (b), and (c) illustrates the aforementioned field distributions 
at $\lambda_{1} =$ 647 nm, respectively. Similar heatmaps at the IR spectral region $\lambda_{2} =$ 1390 nm are illustrated in Fig. \ref{fig:fig4b} (d), (e) and (f). The dominant electric field |E| is confined at the hotspots of the Au nanodisc with an enhancement factor of $20$ (Fig.  \ref{fig:fig4b} (a)). 
However, Fig. \ref{fig:fig4b} (d) shows approximately $40$ folds field enhancement confined in the SiO$_{2}$ region with less dominant fields in the lower edges of the Au nanodisc. Similarly, the magnetic field |H| distribution shows different characteristics of both resonances as well. Thus, the visible band is confined in the Au nanodisc whilst the field in the IR region is confined in the spacer region. 

The current density (J) at $\lambda_{1}$ (Fig. \ref{fig:fig4b} (c)) and at $\lambda_{2}$ (Fig. \ref{fig:fig4b} (f)), reveals the nature of the MIMS structure. It is evident that the current rotates in a loop within Au nanodisc at the visible region. However, in the IR spectral region, we observed antiparallel currents that are excited in the top Au nanodisc and bottom Au film due to the magnetic resonance. The magnetic resonance generates a current in a loop due to the magnetic moment \cite{doi:10.1021/nl9041033}. This leads to strong field enhancement in the spacer region and thereby minimizes the amount of reflected beam in this spectral region. Thus, the electric and magnetic dipole responses of the proposed structure provide means to tune its effective optical properties. By controlling the effective optical properties of the MA, one can tune the effective permittivity and permeability of the structure to impedance match to the superstrate and thereby minimizing the reflective capability of the structure.

These results unveil that the spectral resonance in the visible region appears due to an incident light coupling with the PSP mode of the Au nanodisc and the second band in the IR region corresponds to the existence of the LSP mode. Note that similar absorption peaks in the visible and IR region have also been revealed theoretically by other authors to be linked to the excitation of PSP and LSP modes, respectively \cite{ma13225140, doi:10.1021/nl9041033, Daniel2020, doi:10.1063/1.4878459}.

\subsection{Parametric sweep of MIMS}
To corroborate these mode assignments, we examine the effects of the structural parameters of the MIMS structure. 
\begin{figure}[!ht]
\includegraphics[width=\linewidth]{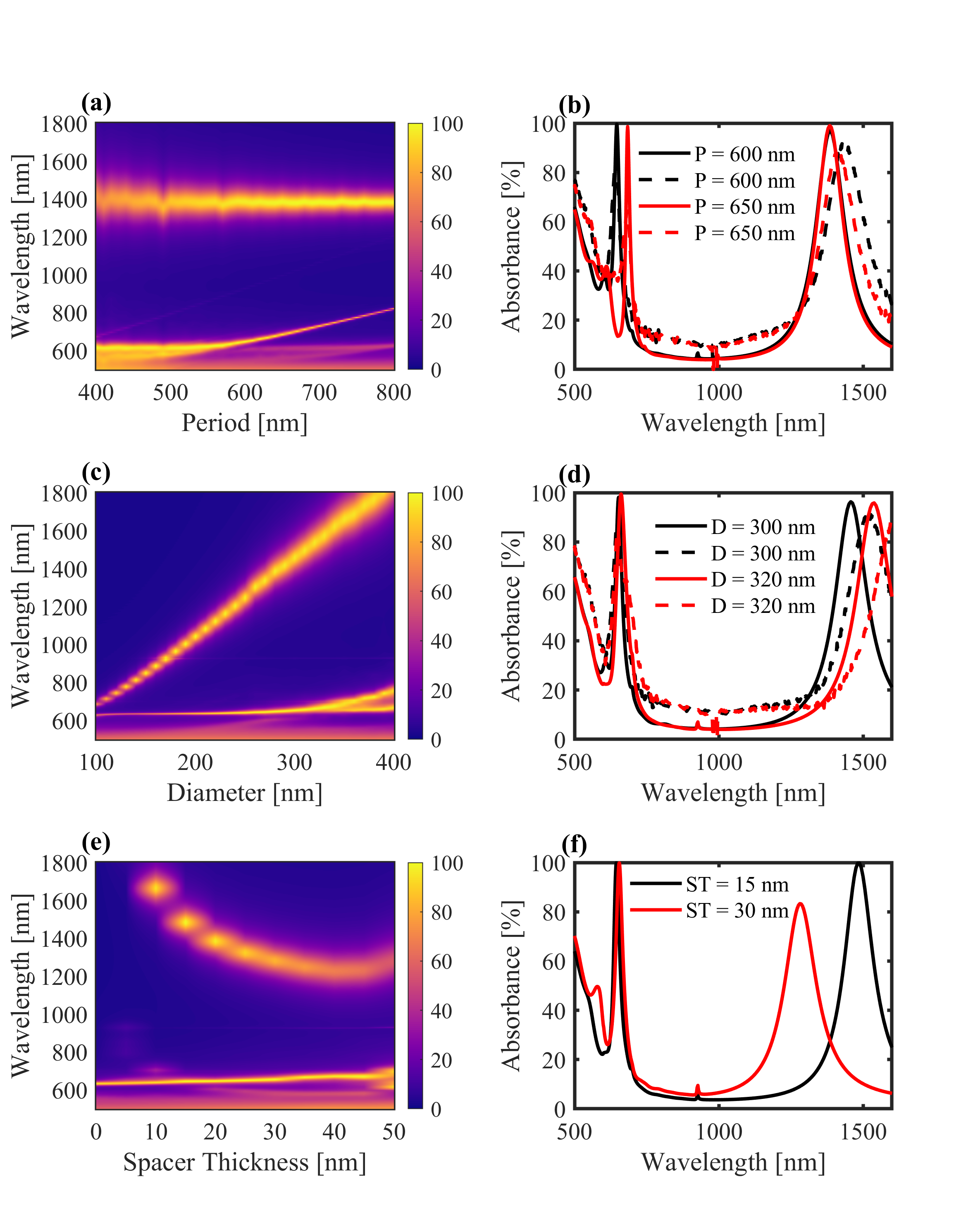}
\caption{\label{fig:fig3} (a) The absorbance as a function of wavelength $\lambda$ and period $P$, and (b) the corresponding selected periods $P$ with experimental results superimposed as dash lines. (c) and (d) Similar plots for diameter $D$ of the top Au nanoarrays as well as spacer thickness $ST$ are shown in (e) and (f).}
\end{figure}
Figure \ref{fig:fig3} illustrates the spectral absorbance  of the MIMS structure as a function of structural parametric sweep of (a) period $P$, (c) diameter $D$, and (e) spacer thickness $ST$. ($A$). Its corresponding 2D plots are shown in (b),  (d), and (f) for the period $P$, diameter $D$, and spacer thickness  $ST$, respectively. The dashed lines superimposed on 
(b) and (d) correspond to the experimentally measured results. It is evident that by changing the periodicity of the nano-patterned Au-array of the MIMS structure, the visible peak absorbance redshifts. This illustrates the link between the absorbance peak excited at $\lambda_{1}$ and the periodicity of the top Au nanoarrays. Thus, the periodic nature of the top nano-patterned Au array induces additional momentum $G=\left(2 \pi / P \sqrt{m^{2}+n^{2}}\right)$, which couples an incident optical field into PSPs mode \cite{ma13225140, ma11030458}. $P$ is the periodicity and $(m, n)$ are the grating orders of the Au nanoarray along both lateral directions. From theoretical formulations, it is evident that the visible peak absorbance corresponds to PSP $(m=1, n=0)$ resonance mode on the top Au nanoarray with air (i.e., refractive index $n = 1$), expressed as,
\begin{equation}
 \lambda_{\mathrm{PSP}} = \frac{P}{\sqrt{m^{2}+n^{2}}} \sqrt{\frac{\epsilon_{m} \epsilon_{d}}{\epsilon_{m}+\epsilon_{d}}}   
\end{equation}
where $\epsilon_{m}$ and $\epsilon_{d}$ are the permittivity of the metal Au and the dielectric surrounding medium, respectively \cite{ma13225140}. From the above formulation, the calculated peak resonance wavelength at the visible wavelength for $(1,0)$ diffracted order (i.e., first-order diffraction mode) is 625 nm, which matches closely to the numerically calculated resonance wavelength ($\lambda_{1}$ = 647 nm) as well as experimental results. This also supports the PSP mode attributed to the peak resonance in the visible region and corroborates the impedance condition (i.e., matching impedance of the structure to the impedance of the surrounding medium) \cite{ma13225140, Xu:20, ZHANG2017216}. 

Moreover, by numerically varying the diameter $D$ of the top Au nanoarray, we determined a shift in the peak resonance in the IR region and confirmed it experimentally. The dependence of the peak resonance on the dimension of the Au nano-arrays as shown in Fig. \ref{fig:fig3} (c) and (d) correlates with the excitation of LSP mode. Thus, the collective excitation of conduction electrons in metallic nanoparticles in response to an incident optical ﬁeld, which is dependent on the size, shape, and surrounding dielectric environment of the Au nanoarrays\cite{doi:10.1021/nl9041033}. 

Similar structural parameter variations for the spacer thickness $ST$ are implemented to comprehend the role of the SiO$_{2}$ sandwiched between the two Au layers. We observed a high peak at the IR region with a spacer thickness of \textit{ST = }\textit{15 nm} (Fig. \ref{fig:fig3} (f)) as well as enhancement of the absorption peak to almost unity ($\approx$ 100\%) as compared with \textit{ST = }\textit{30 nm} with an interesting resonance spectral blueshift.

\subsection{Plasmon-enhanced fluorescence application}

\begin{figure}[!ht]
\includegraphics[width=\linewidth]{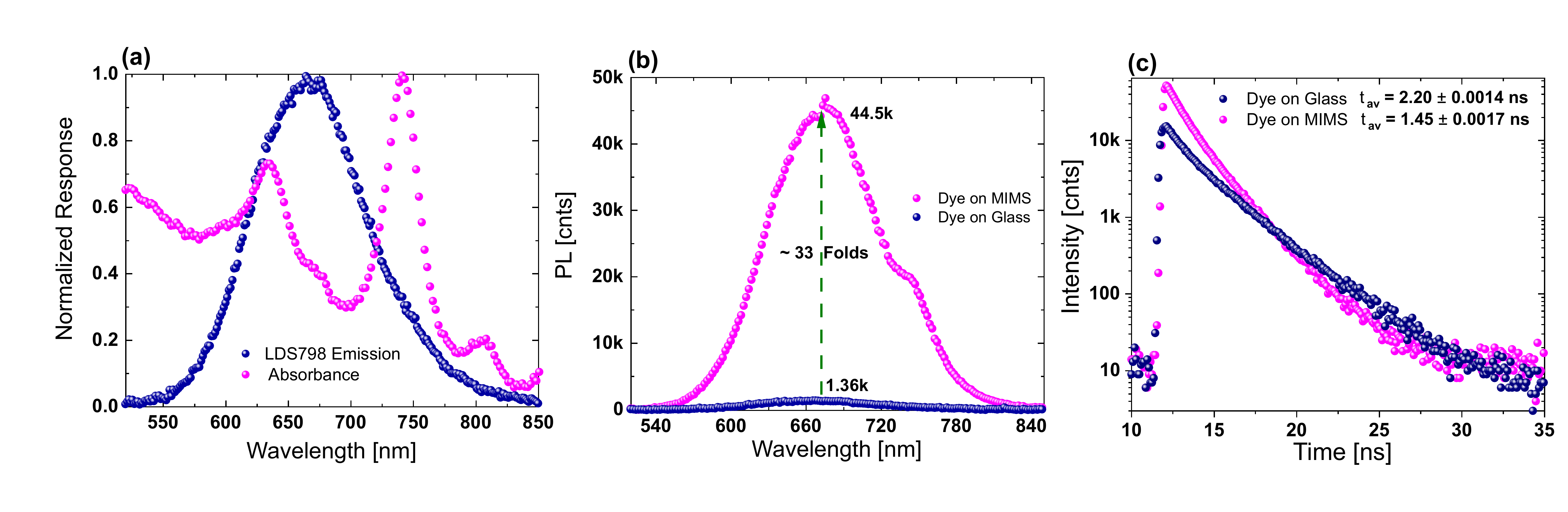}
\caption{\label{fig:fig6} (a) Spectral PSP absorption peak (magenta line) of the nano-patterned MIMS structure with PMMA deposited on it, and superimposed on it (blue line) is the emission spectrum of the LDS798 dye molecules embedded within PMMA. (b) shows the PL response of the dye molecules on the MIMS structure (magenta line) and glass (blue line). (c) shows the corresponding lifetime response of the dye molecules on the MIMS (magenta line) and glass (blue line), respectively.}
\end{figure}

Up until now, we have endeavored to comprehend the plasmon modes that are excited in the MIMS structure and the effect of different surrounding mediums on its spectral absorbance.
To extend this discussion on fluorescence enhancement of dye molecules, we incorporated LDS798 dye molecules into a polymer matrix film (PMMA) to study their emission enhancement capabilities when deposited on top of the MIMS structure. Here, we focused on the degenerate PSP mode of the MIMS which exhibits resonance absorption bands close to the emission wavelength of the dye molecules. Figure \ref{fig:fig6} (a) illustrates the normalized absorbance response of the MIMS structure and the spectral emission of the dye molecules (LDS798) embedded within PMMA. The emission peak of the dye molecule is identified to be at 670 nm and the corresponding degenerate multi-spectral absorption peaks are at $\approx$ 647 nm (weak absorption peak) and  $\approx$ 750 nm (almost unity absorption peak), respectively. These spectral bands of the MIMS overlap with the emission of the dye at different spectral ranges as shown in Fig. \ref{fig:fig6} (a) which could potentially increase the excitation and emission rate of dye molecules (i.e., by opening additional electron states of the fluorophore). To confirm these effects, we experimentally characterized the photoluminescence (PL) response of the dye molecules coupled to the MIMS structure and compared this with a reference dye molecule spin-coated on a glass substrate. As shown in Fig. \ref{fig:fig6} (b), we see high-intensity count rates in the MIMS case (44.5k counts) as compared to the reference sample (1.36k counts) which corresponds to $\approx$ 33 fold emission enhancement. Also, we identified a spectral shift (plateau) in the PL response of the dye molecules coupled to the MIMS structure at the spectral range close to the almost unity absorption peak at $\approx$ 750 nm. Furthermore, we measured the transient fluorescence lifetime values of the dye molecule-PMMA complex coupled with the MIMS structure and compared it to the reference sample as shown in Fig. \ref{fig:fig6} (c). We obtained an average lifetime values of $\tau_{\text{av}} =$ 2.20 $\pm$ 0.0014 $ns$ and $\tau_{\text{av}} =$ 1.45 $\pm$ 0.0017 $ns$ based on a bi-exponential fit of the time-correlated single-photon counting (TCSPC) data for both the reference sample (dye on glass) and the dye molecules deposited on the MIMS structure. Evidently, we observe a shortening of the average lifetime in the MIMS case as compared to the dye on glass. This is due to the strong local electric field generated by the degenerate PSP mode of the MIMS structure and its coupling with the dye molecules. As such, the nonradiative channels of the MIMS structure influence the emission enhancement of the dye molecules and thereby lead to a shortening in the average lifetime values.

\section{Conclusion}

In summary, a robust and feasible dual-band perfect absorber in the visible and IR region based on MIMS structure is proposed and demonstrated. The two optical resonances arise from the PSP and LSP modes in the visible and the IR region, respectively. The performed simulations of the electromagnetic field distribution indicate the strong magnetic field in the spacer region leads to nearly perfect light absorption. Furthermore, the spectral absorbance of the MIMS nanostructure was numerically and experimentally analyzed as a function of the structural parameters to optimize the optical response and corroborate the impedance matching of the structure for the PSP mode. Finally, we exploit the plasmon-enhanced fluorescence capability of the proposed structure by coupling the PSP mode with the emission of the dye molecules. The results presented in this work show the resonant properties of the MIMS structure combine the advantages of metal-insulator-metal (MIM) “perfect absorbers”  with those of plasmonic nanoparticles, enabling the generation of strong enhancements of the electric field and controlling the outcoupling of nonradiative channels.

\begin{backmatter}
\bmsection{Funding}
Academy of Finland Flagship Programme, (PREIN), (320165). 

\bmsection{Acknowledgments}
I.I acknowledge Optica for the Optica Foundation Amplify Scholarship and SPIE for the SPIE Optics and Photonics Education Scholarship. C.R.F acknowledges the European Union’s Horizon 2020 research and innovation programme under the Marie Skłodowska-Curie grant agreement Nº 895369. 
\bmsection{Disclosures}
The authors declare no conflicts of interest.
\bmsection{Data Availability Statement}
Data underlying the results presented in this paper are not publicly available at this time but may be obtained from the authors upon reasonable request.
\end{backmatter}
\bibliography{sample}

\end{document}